\documentclass[11pt]{article}
\usepackage{amsfonts, amssymb,amsmath,epsfig}
 \normalsize
\usepackage{amssymb,amsmath,amsthm,epsfig,amscd,enumerate,eucal,indentfirst}
\newtheorem{theo}{Theorem}[section]

\newtheorem{cor}[theo]{Corollary}

\newtheorem{propo}[theo]{Proposition}

\theoremstyle{definition}
\newtheorem{nb}[theo]{Remark}

\newenvironment{preuve}{\noindent {\bf Proof
:}}{\hfill$\blacksquare$\bigskip}
\def\dt{\mu}
\def\f{\psi}

\def\l{\lambda}
\def\k{\kappa}

\def\a{\alpha}
\def\t{\theta}
\def\b{\beta}

\def \ex {\exp \{-i \xi \cdot \v \}}
\def \exh {\exp \{-i \xi \cdot \vh \}}
\def\Q {\mathcal{Q}}
\def\Qb {\mathcal{Q}_B}
\def \Qh {\widehat{{\mathcal{Q}(M)}}(\xi)}
\def\QF {\mathcal{Q}_{\mathrm{FP}}}
\def\Qe {\mathcal{Q}_{\delta}}

\def\pt{\partial t }
\def\R{{\mathbb R}^3}
\def \S {{\mathbb S}^2}
\def \q {\mathbf{q}}
\def \n {\mathbf{n}}
\def \qn {|\q \cdot \n|}
\def \d {\mathrm{d}}
\def \u {\mathbf{u}_1}
\def \v {\mathbf{v}}
\def \vh {\mathbf{v^{\star}}}
\def \vb {\mathbf{v_{\star}}}
\def \w {\mathbf{w}}
\def \wh {\mathbf{w^{\star}}}
\def \wb {\mathbf{w_{\star}}}
\def\z{2\a (1-\b)}
\def \Itt {\int_{\R \times \R \times \S}}
\def \It {\int_{\R \times \S}}
\def \IS {\int_{\S}}
\def \IR {\int_{\R}}
\def \IRR {\int_{\R \times \R}}
\def \inrt {\int_0^{\infty}\d t}
\def\ep{\epsilon}
\def\cn{\cdot \n}
\def\fd{f_{\var}(\v,t)}
\def\gd{f(\v)}
\def \ft{f(\v,t)}
\def\dpi {\dfrac{1}{2\pi}}
\def\e2{\epsilon^2}
\def\var{\delta}
\newcommand{\bq}{\begin{equation}}
\newcommand{\eq}{\end{equation}}
\newcommand{\bqs}{\begin{equation*}}
\newcommand{\eqs}{\end{equation*}}
\newcommand{\bqm}{\begin{multline*}}
\newcommand{\eqm}{\end{multline*}}

\def\bqa{\begin{eqnarray}}
\def\eqa{\end{eqnarray}}
\def\bd{\begin{displaymath}}
\def\ed{\end{displaymath}}

\numberwithin{equation}{section}

\title{\bf The dissipative linear Boltzmann equation \\ for
hard--spheres}

\title{\bf The dissipative linear Boltzmann equation \\ for
hard spheres}

\author{{\bf Bertrand Lods and Giuseppe Toscani} \\ \normalsize
Dipartimento di Matematica, Universit\'a di Pavia,\\
\normalsize Via Ferrata, 1, 27100 Pavia,
Italy.\\
 \small {\bf e-mail:} {\tt lods@dimat.unipv.it}, {\tt
toscani@dimat.unipv.it}}

\date{}

\begin{document}
\bibliographystyle{plain}
\maketitle

\begin{abstract}
\noindent We prove the existence and uniqueness of an equilibrium state with unit mass to the dissipative linear
Boltzmann equation with hard--spheres collision kernel describing inelastic interactions of a gas particles with a
fixed background. The equilibrium state is a universal Maxwellian distribution function with the same velocity as
field particles and  with a non--zero temperature lower than the background one, which depends on the details of
the binary collision. Thanks to the $H$-theorem we then prove strong convergence of the solution to the Boltzmann
equation towards the equilibrium.
\end{abstract}

{\small \noindent {\bf Key words.} Granular gases, equilibrium
state, linear Fokker--Planck equation, trend to equilibrium.



\section{Introduction}

The present paper deals with the linear dissipative Boltzmann equation for hard spheres interactions, and solves
some questions left open in the very recent work  by Toscani and Spiga \cite{spiga}, devoted to the investigation
of a linear dissipative Boltzmann equation for Maxwell molecules.

In the last years,  the study of kinetic models for granular flows received a significant interest. The largest part of this
work deals with kinetic {\it nonlinear models} based upon generalizations of the Boltzmann--Enskog equation. We refer the
interested reader to the review articles \cite{cerci1, goldh, goldh1}. However, most of the studies refer to inelastic Maxwell
particles, both for the driven case \cite{bobcer, carrillo} or for the free case \cite{bobgam, BC4, BCT}. Such
(pseudo--)Maxwellian models enjoy nice mathematical simplifications and lead to exact analytical results \cite{ernst, brito}
(see also the recent developments on the inelastic Kac model \cite{toscani, ada}). At present, only a few papers consider real
interactions among grains, which are well described by the inelastic hard--spheres models \cite{bobpanf, gampanf}.

Despite their importance for practical applications, linear equations for dissipative models have been much less
addressed. To our knowledge the only progresses on the matter are those of the afore--mentioned paper \cite{spiga}
and of R. Pettersson \cite{petter}. Linear models describe the time evolution of the distribution function
$f(x,\v,t)$ of particles of masses $m$ (representing the granular gas) colliding {\it inelastically} with particles
with masses $m_1$ of a fixed background. Throughout this paper, the subscript $(1)$ will be addressed to the fixed
field particles whose distribution function is known and is assumed to be a normalized Maxwellian $M_1$ with given
mass velocity and temperature. Note that, the grains being cohesionless, long--range interactions of any kind are
irrelevant. Thus, the only model with real physical interest is the \textit{hard--spheres model}. As introduced in
\cite{spiga}, the evolution of $f(x,\v,t)$ is given by
\bq\label{bolt} \dfrac{\partial f}{\partial t}(\v,t)+\v \cdot
\nabla_x f(x,\v,t)=\dfrac{1}{2\pi\lambda} \It \qn
\left[\dfrac{1}{\e2}f(\vb)M_1(\wb)-f(\v)M_1(\w)\right]\d\w \d\n.
\eq
Here $\lambda$ denotes the constant mean free path, $\q$ is the relative velocity, $\q=\v-\w.$ The velocities
$(\vb,\wb)$ are the pre--collisional velocities of the so--called inverse collision, which results in $(\v,\w)$ as
post-collisional velocities. In the granular setting, the most important feature of the collision mechanism is its
{\it inelastic character} which induces that (generally)  \textit{the total kinetic energy is dissipated}. The
constant parameter $0< \ep <1$ is called the restitution coefficient and measures the inelasticity of the
collisions. Whenever $\ep=1$ we recover the usual linear Boltzmann equation (see Section \ref{bol} for
details).\medskip

One of the main feature of this paper is to prove the existence
and uniqueness of the (homogeneous) equilibrium state of equation
\eqref{bolt}. Precisely, we exhibit a (non-trivial) distribution
function $M$ (depending on the velocity only) such that
$$\Q(M)=0$$
where $\Q$ denotes the right-hand side operator of \eqref{bolt}.
When $\ep=1$, this question is trivial since the conservation of
momentum and energy implies
$$\widetilde{M}(\v)M_1(\w)=\widetilde{M}(\vb)M_1(\wb)$$
where $\widetilde{M}$ stands for the Maxwellian distribution with
same drift velocity and temperature as $M_1$ but corresponding to
particles of masses $m$ (see Section \ref{bol} for more details).
Then, one sees immediately that the integrand of $\Q$ vanishes for
$f=\widetilde{M}.$ Clearly, for $0< \ep < 1$ this is no more the
case and it appears difficult to determine explicitly the
(eventual) equilibrium state. Actually, the two following
questions are far for being trivial:
\begin{itemize}
\item Does an equilibrium state exist?
\item If it does, is it given by some suitable Maxwellian distribution?
\end{itemize}
In this paper, we answer positively to both questions showing that
the equilibrium state is a Maxwellian distribution with the same
mean velocity as $M_1$ and with a universal non-zero temperature
lower than the given background temperature. Moreover, this
equilibrium state is unique, provided its mass is prescribed. This
Maxwellian distribution coincide with the one derived in the
pseudo--Maxwellian case \cite{spiga}. Actually, it is also
possible to show that, as in the {\it non--dissipative} case, the
equilibrium state is universal in the sense that its temperature
does not depend on the collision kernel, but only on the
inelasticity parameter $\ep$ which determines the details of the
binary interaction. Let us mention here that our results answer to
some open questions from \cite{spiga} and complete the study of
\cite{petter} where the existence of an equilibrium state was used
as an assumption for some of the results. \medskip

The two previous questions, as well as the problem of the rate of
convergence towards equilibrium, have been recently addressed in
\cite{spiga} for the \textit{pseudo-Maxwellian approximation}.
This pseudo-Maxwellian approximation consists in replacing the
relative velocity $\q$ appearing in the collision kernel $\qn$ of
$\Q$ by the unit vector in the direction of $\q$. The
pseudo-Maxwellian model enjoys in particular two fundamental
properties. First, the associated \textit{moment equations are
closed} with respect to the moments of the distribution function.
Hence, it is possible to derive the time evolution of the drift
velocity $\mathbf{u}(t)$ and the temperature $T(t)$ of $f(\v,t)$
and to predict the mass velocity and temperature of the eventual
equilibrium state. Moreover, as pointed out by A. Bobylev
\cite{bob}, Maxwell models lend themselves to a convenient Fourier
analysis. These two important properties enabled to determine the
Maxwellian equilibrium state for the pseudo-Maxwellian model and
to prove also exponential convergence of the solution to
\eqref{bolt} towards the equilibrium (in the homogeneous setting).

Unfortunately, these two tools \textit{do not apply} to the hard--spheres model \eqref{bolt} and one has to proceed in a
different way. The main problem is actually to predict what should be the eventual steady-state. To this aim, we derive
\textit{formally} a linear Fokker--Planck equation which is naturally associated to the dissipative Boltzmann model
\eqref{bolt} (Section \ref{graz}) through the asymptotic of the grazing collisions (see e.g. \cite[Chapter II.9]{cerci}).
There is  a noticeable amount of results on the matter for the elastic (nonlinear) Boltzmann equation. We mention here the
papers \cite{villa1, villa2, degond, desvill} which enlighten  the connection between the nonlinear Boltzmann equation and
the Landau--Fokker--Planck equation and the works \cite{ricci, goudth} which describe the same procedure for linear problems.
We emphasize the fact that the main goal of this analysis is  not to prove rigorously any kind of asymptotic procedure, even
if it is possible to make
 our results rigorous (see Appendix). The method of Section 3 must only be viewed as a formal (but efficient) way
to find a suitable approximation of the collision operator $\Q$ which maintains the equilibrium distribution. To our knowledge,
it is the first time (in kinetic theory) that such a limiting process is performed within this scope. The utility of this
procedure comes out from the fact that the equilibrium state of the Fokker--Planck operator is immediate to obtain. It is  a
Maxwellian distribution which appears then as the candidate to be the stationary solution of \eqref{bolt}. The main problem is
then to show that this Maxwellian is effectively a steady state for $\Q.$ This will be done by means of Fourier transform
arguments.
\medskip

A second important task addressed in this paper is the large--time
behavior of the solution to the space homogeneous version of
\eqref{bolt}. Actually, once the existence and uniqueness of an
equilibrium state established, the question of trend towards this
equilibrium is of primary importance in kinetic theory. In this
paper, we are able to show a strong $L^1$--convergence result to
the Maxwellian steady--state. As in \cite{petter}, our result is
based upon weak--compactness arguments by means of the so--called
$H$-theorem and some estimates on the second moment of the
solution to \eqref{bolt}. \medskip

Let us explain now in some details the organization of the paper. In Section 2, we describe briefly the dissipative Boltzmann
linear model and its properties. In Section 3 we deal with the asymptotics of the grazing collision, from which we derive the
Fokker--Planck  approximation of $\Q$ which help us to identify the (possible) equilibrium state. Then in Section 4 we show
that the Maxwellian obtained through the grazing collisions procedure is really a stationary solution to \eqref{bolt}. Thanks
to the Boltzmann $H$--theorem we then prove in Section 5 that the equilibrium state is unique (provided its mass is
prescribed) and that the solution to \eqref{bolt} converges (in the strong $L^1$--sense) towards the equilibrium. Finally, we
end this paper by presenting some open questions and perspectives.

%
\section{The dissipative linear Boltzmann equation}\label{bol}

As briefly described in the introduction,  in this paper we are concerned with the evolution of the distribution function
$f(\v,t)$ of granular gas particles with masses $m$ which undergo inelastic collisions with the field particles (of masses
$m_1$) of a fixed background. The background is supposed to be at thermodynamical equilibrium with given temperature $T_1$
and given mass velocity $\u$, i. e. its distribution function is the  {\it normalized Maxwellian}
\bqs 
M_1(\v)=\left(\dfrac{m_1}{2\pi T_1}\right)^{3/2}\exp
\{-\dfrac{m_1(\v-\u)^2}{2T_1}\} \qquad \qquad \v \in \R.\eqs
 The main feature of the binary dissipative
 collisions is that part of the normal relative
velocity is lost in the interaction, so that
\begin{equation}\label{coef}
(\vh-\wh) \cn = - \ep (\v-\w)\cn ,
\end{equation}
where $\n \in \S$ is the unit vector in the direction of impact, $(\v,\w)$ stand for the velocities before impact
whereas $(\vh,\wh)$ denote the post--collisional velocities.  The so-called (constant) \textit{restitution
coefficient} $\ep$ is such that $0 < \ep < 1$, the case $\ep=1$ corresponding to elastic collisions. Thanks to
\eqref{coef} and assuming the conservation of momentum
$$m\vh+m_1\wh=m\v+m_1\w$$ one then finds the collision
mechanism
\begin{equation}\label{mechah}\begin{cases}
\vh=\v-2\alpha(1-\beta) [(\v-\w) \cn] \n\\
\wh=\w+2(1-\alpha)(1-\beta) [(\v-\w) \cn] \n,
\end{cases}\end{equation}
where $\alpha$ is the mass ratio and $\beta$ denotes the inelasticity parameter
$$\alpha=\dfrac{m_1}{m+m_1} \qquad \qquad \qquad \beta=\dfrac{1-\ep}{2}.$$
The parameter $\alpha$ is such that $0 < \alpha < 1$ (we exclude the limiting cases of Lorentz and Rayleigh gases), while the
inelasticity parameter satisfies $0 < \beta < 1/2$. We refer to \cite{gampanf} for a detailed description of the geometry of
the collisions. It is easy to see that system \eqref{mechah} is invertible and provides the pre--collisional velocities of
the so-called inverse collisions, resulting in $(\v,\w)$ as post-collisional velocities
\bqs \label{mechab}
\begin{cases}
\vb=\v-2\alpha\dfrac{1-\beta}{1-2\beta} [(\v-\w) \cn] \n\\
\wb=\w+2(1-\alpha)\dfrac{1-\beta}{1-2\beta} [(\v-\w) \cn] \n.
\end{cases}\eqs
In contrast to the elastic case ($\ep=1$), the binary collision  dissipates part of the kinetic energy
$$m |\vh|^2 + m_1 |\wh|^2 -(m |\v|^2 + m_1 |\w|^2) = -4 \dfrac{mm_1}{m+m_1}\b(1-\b) \qn ^2 \leq 0.$$
In space homogeneous conditions, upon using $\l$ as a time scale, equation
  \eqref{bolt} can be re-written in the dimensionless form
\begin{equation}\label{bolt1}
\dfrac{\partial f}{\partial t}(\v,t)=\dfrac{1}{2\pi} \It \qn
\left[\dfrac{1}{\e2}f(\vb)M_1(\wb)-f(\v)M_1(\w)\right]\d\w \d\n.
\end{equation}
The factor $\ep^{-2}$ in the gain term above appears respectively
from the Jacobian of the transformation $\d\vb \d\wb$ into
$\d\v\d\w$ and from the length of the cylinders $|\q^{\star}
\cn|=\ep \qn$ (see \cite{cerci1} for details). Let $\Q$ be the
(dissipative) linear Boltzmann collision operator (acting only on
the velocity space)
\bq \label{Q} \Q(f)=\dfrac{1}{2\pi}\It \qn
\left[\dfrac{1}{\e2}f(\vb)M_1(\wb)-f(\v)M_1(\w)\right]\d\w \d\n
\eq
The change of variables $\n \to -\n$ leads to the equivalent expression
\bqs \Q(f)=\dfrac{1}{\pi}\It H(\q \cn /|\q|) \q \cn \left[\dfrac{1}{\e2}f(\vb)M_1(\wb)-f(\v)M_1(\w)\right]\d\w \d\n ,
\eqs
where $H(\cdot)$ is the Heavyside step function. We can also define the collision operator by its action on the {\it
observables}. Precisely, for any regular test--function $\psi(\v)$
\bq \label{weak} \langle \psi, \Q(f) \rangle=\dpi \Itt \qn
f(\v)M_1(\w)[\psi(\vh)-\psi(\v)]\d\v\d\w\d\n.\eq
Clearly, $\psi(\v)\equiv 1$ is a collision invariant (mass
conservation) whereas, in contrast to the elastic case,
$\psi(\v)=\v$ and $\psi(\v)=\v^2$ are not (dispersion of kinetic
energy). Note that an important feature of the hard--spheres model
is that (even in the elastic case) the moments equations of
$\Q(f)$ are not closed with respect to the ones of $f$.
%
%
\section{The Fokker--Planck approximation}\label{graz}

The main goal of this section is the {\it formal} derivation of a
linear Fokker--Planck equation, obtained from \eqref{bolt1}
through a kind of grazing collisions asymptotics. We point out
that, our aim in this paper, \textit{is not} to prove rigorously
the convergence of the (re-scaled) dissipative Boltzmann operator
$\Q$ towards the Fokker--Planck operator $\QF$ \eqref{FP} below as
the collisions become grazing. \textit{The limiting process we
perform here must only be seen as an {\it efficient tool to
predict the nature of the equilibrium state} of} $\Q$ (if it
exists). Nevertheless, this approximation result can be made
rigorous and this shall be done in the Appendix. Here we proceed
only  at a formal level. Let us assume that all the collisions
concentrate around
\bq \label{qn} |\q \cn| /|\q| \sim 0.\eq
Consequently, according to \eqref{mechah} one has $|\vh-\v| \sim
0$ and, for any smooth function $\f$, one can perform a Taylor
expansion of $\f(\vh)$ around $\v$ leading, at the second order,
to:
\bq \begin{split}\label{taylor} \f(\vh)&=\f(\v)+\nabla_\v \f(\v)
\cdot (\vh-\v) + \frac{1}{2} \mathbb{D}^2\f(\v) (\vh-\v) \otimes
(\vh-\v)+ \mathrm{o}(|\vh-\v|^2)\\
&=\f(\v) -2\a (1-\b) (\q \cn) \nabla_\v \f(\v) \cn +\frac{\left( 2\a (1-\b)) \q \cn \right)^2}{2}\mathbb{D}^2\f(\v) \cdot \n
\otimes \n + \mathrm{o}(|\q \cn /|\q| |^2). \end{split} \eq
In (\ref{taylor}) $\mathbb{D}^2\f$ is the Hessian matrix of $\f$. The o$(|\vh-\v|^2)$ term will be neglected in the sequel.
One clearly observes that the expansion \eqref{taylor} is similar to that obtained in the study of elastic collisions between
particles of same masses \cite{goudth}. This property obviously comes from the fact that \eqref{qn} implies also $\v' \sim
\v$ where $\v'$ is the post--collisional velocity in the elastic case. The only difference is that, in the \textit {classical
elastic} theory, the multiplicative constant $2\a (1-\b)$ is taken to be equal to $1/2$. Thus, staying at a formal level,
\textit{ dissipative collisions between particles of unequal masses does not lead to supplementary difficulties.}
\medskip

Let us consider a referential frame with the $x$-axis directed
along $\q$. Then,
 \bq \label{ncos}
\begin{cases}
\n=(\cos \t, \sin \t \cos \xi, \sin \t \sin \xi) \qquad \qquad 0 \leq \t \leq \pi/2, \:0 \leq \xi \leq 2\pi,\\
\cos \t=\dfrac{\qn}{|\q|} \qquad \text{ and } \qquad \d\n=\sin
\t\, \d \t \d \xi.
\end{cases}\eq
Assuming that the collisions concentrate around $\t \sim \pi/2$,
we define
$$b_{\var}(\t)=\dfrac{2}{\pi}\chi_{[\pi/2-\var,\pi/2]}(\t) \qquad \qquad (\delta > 0)$$
and
$$I_{\var}=\int_0^{\pi/2}b_{\var}(\t)\cos^3 \t\, \sin \t \,\d\t.$$
One sees that
\bq \label{ie} I_{\var} \sim \dfrac{\var^4}{2\pi} \qquad \text{ as
} \qquad \var \sim 0.\eq
Consequently, let us define the associated collision kernel
$$B_{\var}(\q,\n)=b_{\var}(\t)\qn ,$$
and denote by $\Qe$ the collision operator obtained by replacing
$\qn$ by $\var^{-4}B_{\var}(\q,\n)$ in \eqref{bolt1}.
\begin{nb}\label{scale} Note that the introduction of the multiplicative factor
$\var^{-4}$ can be seen as a suitable {\it time--scaling} in
\eqref{bolt1} (see Appendix \ref{appen} for further
details).\end{nb}

Our aim is to show that for small values of $\delta$, the (re-scaled) operator $\Qe$ is closed to be a Fokker--Planck
collision operator. Precisely, let us fix $f \in L^1(\R)$ and a smooth test--function $\psi(\v).$ Using $\Qe$ into
\eqref{weak} leads to
\bqs \langle
\psi,\,\Qe(f) \rangle = \dfrac{1}{2\pi \var^{4}} \Itt
B_{\var}(\q,\n) \gd M_1(\w) \left[ \psi(\vh)-\psi(\v)\right] \d \v
\d \w \d \n.\eqs
Now, inserting the expansion \eqref{taylor} in the above expression gives the second order approximation
\begin{multline}\label{tay}
\langle \psi,\,\Qe(f) \rangle
=\mathbf{J}^1_{\var}+\mathbf{J}^2_{\var}=\\
= - \dfrac{\z}{ 2\pi \var^4} \Itt B_{\var}(\q,\n)(\q \cn) \gd
M_1(\w) \nabla_\v
\f(\v) \cn \,\d\v\d\w\d\n\\
+ \dfrac{(\z )^2}{4\pi \var^4} \Itt B_{\var}(\q,\n)\left(\q \cn
\right)^2 \gd M_1(\w) \mathbb{D}^2 \f(\v) \cdot \n \otimes \n
\,\d\v\d\w\d\n.
\end{multline}
To estimate $\mathbf{J}^1_{\var}$, we first compute the integral
with respect to $\d\n$. According to \eqref{ncos}
\bqs
\begin{split} \IS B_{\var}(\q,\n)(\q \cn) \n \d\n&= 2\pi |\q|
\int_{0}^{\pi/2} B_{\var}(\q,\n)(\cos^2 \t,0,0) \sin \t
\, \d\t\\
&= 2\pi |\q|^2 \int_{0}^{\pi/2} b_{\var}(\t)(\cos^3 \t,0,0)\sin \t
\d
\t\\
&= 2 \pi |\q|^2(I_{\var},0,0)= 2\pi I_{\var} |\q| \q
.\end{split}\eqs
Therefore
$$\mathbf{J}^1_{\var}=-I_{\var}\dfrac{\z}{\var^4}\IRR |\q|^2 \gd
M_1(\w) \nabla_\v \f(\v,t) \cdot \dfrac{\q}{|\q|} \,\d\v\d\w .$$
Thanks to \eqref{ie},  the coefficient in front of the above integral converges to $-\a(1-\b)/\pi$  as $\var$ goes to $0$.
Consequently
\bq \label{limj1} \mathbf{J}^1_{\var} \simeq -\frac{\a(1-\b)}{\pi}
\IRR |\q|^2 \gd M_1(\w) \nabla_\v \f(\v,t) \cdot \dfrac{\q}{|\q|}
\,\d\v\d\w.\eq
We proceed in the same way for $\mathbf{J}^2_{\var}.$ One has
first
\bq
\begin{split}
\IS B_{\var}(\q,\n)\left(\q \cn \right)^2 \n \otimes \n \, \d\n
&=|\q|^3 \int_0^{\pi/2} b_{\var}(\t) \cos^3
\t \, \sin \t\,\d \t \int_{0}^{2\pi} n \otimes n \,\d\xi\\
&=2\pi |\q|^3 \int_0^{\pi/2} b_{\var}(\t) \cos^3 \t \, \sin \t\,\mathbf{Diag}[\cos ^2 \t,
\frac{1}{2}\sin^2\t,\frac{1}{2}\sin^2\t] \d \t ,\end{split}\eq
where $\mathbf{Diag}[a_1,a_2,a_3]$ is the diagonal matrix in $\R
\times \R$ whose diagonal entries are $a_i$ $(i=1,2,3).$ Now,
defining
$$K_{\var}=\int_{0}^{\pi/2}b_{\var}(\t)\cos^5 \t\, \sin \t \,\d\t ,$$
one gets
\bqs \IS B_{\var}(\q,\n)\left(\q \cn \right)^2 \n \otimes \n \, \d\n = 2\pi |\q|^3 \mathbf{Diag}[K_{\var},
\frac{1}{2}(I_{\var}-K_{\var}), \frac{1}{2}(I_{\var}-K_{\var})],
\eqs
and
\bqs \mathbf{J}^2_{\var}=\dfrac{(\z )^2}{2 \var^4} \IRR |\q|^3 \gd
M_1(\w)  \mathbb{D}^2\f(\v,t) \cdot \mathbf{Diag}[K_{\var},
\frac{1}{2}(I_{\var}-K_{\var}), \frac{1}{2}(I_{\var}-K_{\var})]
\d\v\d\w. \eqs
Since $K_{\var}$ is negligible with respect to $I_{\var}$ one concludes with the following approximation
\bq \label{limj2} \mathbf{J}^2_{\var} \simeq
\dfrac{\a^2(1-\b)^2}{2\pi} \IRR |\q|^3 \gd M_1(\w)
\mathbb{D}^2\f(\v) \cdot \mathbf{Diag}[0,1,1]\,\d\v\d\w.\eq
For any $\mathbf{z} \in \R$ $(\mathbf{z} \neq 0)$, let $\mathbb{S}(\mathbf{z})$ be the symmetric \textit{matrix}
$$\mathbb{S}(\mathbf{z})=\mathrm{Id}-\dfrac{\mathbf{z} \otimes \mathbf{z}}{|\mathbf{z}|^2},$$
i.e. $\mathbb{S}(\mathbf{z})$ is the projection on the space orthogonal to $\mathbf{z}$. Then, combining \eqref{tay},
\eqref{limj1} and \eqref{limj2}, one obtains  for $\Qe$ the  approximation
\begin{multline*}
\langle \psi,\,\Qe(f) \rangle  \simeq -\frac{\a(1-\b)}{\pi} \IRR
|\q|^2 \gd M_1(\w) \nabla_\v \f(\v) \cdot
\dfrac{\q}{|\q|} \,\d\v\d\w\\
+ \dfrac{\a^2(1-\b)^2}{2\pi} \IRR |\q|^3 \gd M_1(\w)
\mathbb{D}^2\f(\v) \cdot \mathbb{S}(\v-\w) \,\d\v\d\w.
\end{multline*}
Now, straightforward computations, using the fact that
$2|\q|\q=\mathbf{Div}_{\v}(|\v-\w|^3\mathbf{S}(\v-\w)),$ yield
\bqs \langle \psi,\,\Qe(f) \rangle  \simeq
 - \dpi \IR  \d\v \,  \nabla_{\v}\psi(\v)
 \IR |\v-\w|^3 \mathbb{S}(\v-\w)
\left\{\k M_1(\w)\nabla_{\v}f(\v) + (\k -\mu)
f(\v)\nabla_{\w}M_1(\w)\right\}\d\w \eqs
where the parameters $\k$ and $\mu$ are defined respectively by
$$\k=\a^2 (1-\b)^2 ; \qquad
\mu=\alpha (1-\b).$$
Since the above approximation is valid for any arbitrary smooth function $\f$, one sees that, as $\var$ goes to 0, $\Qe$ can
be approximated (up to the constant $1/2\pi$) by the  Fokker--Planck operator
\bqs \QF (g)(\v)= \nabla_{\v} \cdot \IR |\v-\w|^3
\mathbb{S}(\v-\w)\cdot
 \left\{\k M_1(\w)\nabla_{\v}g(\v) + (\k
-\mu) g(\v)\nabla_{\w}M_1(\w)\right\}\d\w. \eqs
Let us write $\QF$ in a nicer way. Using the fact that $\mathbb{S}(\v-\w) \cdot (\v-\w)=0$ and
$$\nabla_{\w}M_1(\w)=-\dfrac{m_1(\w-\u)}{T_1}M_1(\w),$$
one has
\bq \label{FP} \begin{split}
 \QF (g)(\v)&=\k \nabla_{\v} \cdot \IR
|\v-\w|^3M_1(\w) \mathbb{S}(\v-\w)\cdot \left\{\nabla_{\v}g(\v) -
\dfrac{m_1(\k -\mu)}{\kappa T_1}(\v-\u)g(\v)\right\}\d\w \\
&= \k \nabla_{\v} \cdot \left [\mathbb{A}(\v) \cdot \left\{ \nabla_{\v}g(\v) + \dfrac{m_1(\mu -\k)}{\kappa T_1}(\v-\u)g(\v)
\right \} \right ],\end{split} \eq
where $\mathbb{A}(\v)$ denotes the invertible matrix
$$\mathbb{A}(\v)=\IR
|\v-\w|^3M_1(\w) \mathbb{S}(\v-\w)\,\d\w.$$
\begin{nb} In a different spirit, J. J. Brey \textit{et al.}
\cite{brey} derived a linear Fokker--Planck equation from
\eqref{bolt1} in the limit of small mass ratio $(\alpha \to 0)$.
We also point out that it is possible to consider (see e. g.
\cite{toscan}) the quasi--elastic approximation of $\Q$ assuming
that $\b \ll 1.$ We adopt here the grazing collisions asymptotic
since it preserves the parameters $\a$ and $\b$ (and therefore the
elasticity) modifying only the geometry of the collisions. Note
however that the existence of an equilibrium state for the linear
quasi--elastic approximation of $\Q$ is an open problem to our
knowledge.
\end{nb}

Now, one recognizes immediately that the above procedure preserves
the (eventual) equilibrium state. Precisely, if $F$ is an
equilibrium state of $\Q_{\var}$, then $\QF(F)=0.$ This strongly
suggests that one has to select the candidate for being an
equilibrium state of $\Q$ as the one of $\QF$. Of course, the
interest of the above procedure lies in the fact that this latter
is easy to exhibit. Indeed, it is obvious from \eqref{FP} that the
unique solution with unit mass to $\QF(g)=0$ is given by a
Maxwellian distribution with drift velocity $\u$ and temperature
$$\dfrac{T^{\#}}{m}=\left\{\dfrac{m_1(\mu-\k)}{\kappa T_1}\right\}^{-1}.$$
This suggests the following dichotomy:
\begin{itemize}
\item Either $\Q(f)=0$ has no non-trivial solution.
\item Either the unique solution to $\Q(f)=0$ with unit mass is
the Maxwellian:
\bq \label{maxw} M(\v)=\left(\dfrac{m}{2\pi
T^{\#}}\right)^{3/2}\exp \{-\dfrac{m(\v-\u)^2}{2T^{\#}}\} \qquad
\qquad \v \in \R \eq
where
\bq \label{temp} T^{\#}=\dfrac{(1-\a)(1-\b)}{1-\a(1-\b)}T_1.\eq
\end{itemize}

At this point, we remark that the above Maxwellian distribution is
exactly the equilibrium state found in \cite{spiga} in the study
of the pseudo-Maxwellian approximation of \eqref{bolt1}. This
supports our belief that $M$ is indeed the steady state of $\Q$
and that, moreover, it is also a {\it universal} stationary
solution (independent of the collision kernel) as it occurs in the
elastic case. This will be proved rigorously in the following
section.

\section{Is the Maxwellian the equilibrium state ?}\label{equili}

The problem of finding the equilibrium state of the linear
Boltzmann equation \eqref{bolt1} has now been reduced to determine
whether $\Q(M) \equiv 0$ or not, where
$$M(\v)=\left(\dfrac{m}{2\pi T^{\#}}\right)^{3/2}\exp
\{-\dfrac{m(\v-\u)^2}{2T^{\#}}\},  \qquad  \v \in \R,$$
with $T^{\#}=\dfrac{(1-\a)(1-\b)}{1-\a(1-\b)}T_1.$ Surprisingly, apart for the peculiar case of the 1D-model, we have not
been able to prove by direct computation that
$$\Q(M)(\v)=0, \qquad \v \in \R.$$
Actually, we prove this result through Fourier analysis, i. e. we
show that
$$\Qh =0 , \qquad  \xi \in \R,$$
where $\Qh$ denotes the Fourier transform of $\Q(M)$. By
\eqref{weak}, it is given by
\bqs \Qh = \Itt \qn M(\v)M_1(\w)[ \exh -\ex ]\d\v\d\w\d\n. \eqs
It is immediate to show that, up to a translation of the referential frame, one can assume
$$\u=0.$$
For the sake of simplicity, we introduce the  parameter
$$C= \left(\dfrac{m\,m_1}{2T_1\,T^{\#}}\right)^{3/2}$$
and we recall that $\mu=\a(1-\b).$
Then,
\begin{multline*} \Qh = \IRR |\q| M(\v)M_1(\w) \ex \d\v\d\w \times \\
\times \IS |\frac{\q}{|\q|} \cn |\,\left(\: \exp \{ 2\,i\,\dt\,(\q
\cn)\,(\xi \cn)\} -1\: \right) \d\n. \end{multline*}
Now, the \textit{key point of our computations} is the identity
\bqs\begin{split} M(\v)M_1(\w)&=C\,\exp \left\{ -\frac{m_1}{2 \dt
T_1}\left[ \dt \w^2 +(1-\dt) \v^2 \right] \right\}\\
&=C\,\exp \left\{ -\frac{m_1}{2 \dt T_1} \left[ \dt(1-\dt)\q^2 + (\v-\dt \q)^2\right]\right\},
 \end{split}\eqs
that follows from the equality $\dfrac{m}{T^{\#}}=-(\dt-1) \dfrac{m_1}{\dt T_1}.$ Then,
\begin{multline*} \Qh = C \IRR |\q| \exp \left\{ -\frac{m_1}{2 \dt T_1} \left[ \dt(1-\dt)\q^2 +
(\v-\dt \q)^2\right]\right\}\ex \d\v\d\w \times \\
\times  \IS |\frac{\q}{|\q|} \cn |\,\left(\: \exp \{
2\,i\,\dt\,(\q \cn)\,(\xi \cn)\} -1\: \right) \d\n.
\end{multline*}
The change of variables $(\v,\w) \to (\v,\q)$ yields
\begin{multline*}
\Qh = C \IR |\q| \exp \{-\frac{m_1}{2 \dt T_1} \dt(1-\dt)\q^2\}
\d\q \times \\
\times \IR \exp \{ -\frac{m_1}{2 \dt T_1} (\v-\dt \q)^2 \} \ex \d\v \times \\
\times \IS |\frac{\q}{|\q|} \cn |\,\left(\: \exp \{ 2\,i\,\dt\,(\q
\cn)\,(\xi \cn)\} -1\: \right) \d\n.
\end{multline*}
Performing first the second integral leads to
\begin{multline*}
\Qh = C \exp \{- \frac{\dt T_1}{2 m_1} \xi^2\} \IR |\q| \exp
\{-\frac{m_1}{2 T_1} (1-\dt)\q^2\} \exp \{ -i \dt \q \cdot \xi\}
\d\q \times \\
\times \IS |\frac{\q}{|\q|} \cn |\,\left(\: \exp \{ 2\,i\,\dt\,(\q \cn)\,(\xi \cn)\} -1\: \right) \d\n,
\end{multline*}
where we used the fact that the Fourier transform of the Gaussian
$\exp\{- (\v-\mathbf{u})^2/2 \Theta\}$ is equal to
$ C_{\Theta} \exp \{ -i \mathbf{u} \cdot \xi - \Theta/2 \xi^2 \}$
for any $\Theta > 0$ and $\mathbf{u} \in \R$ (here $\mathbf{u}=\dt
\q$ and $\Theta=m_1/\dt T_1$) where $C_{\Theta}$ is a
multiplicative constant depending only on $\Theta.$ Now, as
pointed out first by A. Bobylev \cite{bob}, the inner integral on
the unit sphere is an isotropic function of the vectors $\xi$ and
$\q$ and is therefore equal to
$$\IS |\xi  \cn / |\xi| |\,\left(\: \exp \{ 2\,i\,\dt\,(\q
\cn)\,(\xi \cn)\} -1\: \right) \d\n.$$
Consequently,
\begin{multline*}
\Qh = C \exp \{- \frac{\dt T_1}{2 m_1} \xi^2\} \IR |\q| \exp
\{-\frac{m_1}{2 T_1} (1-\dt)\q^2\} \d\q \times \\
\times \IS |\xi \cn /|\xi||\,\left(\: \exp \{ -i\,\dt\,(\q \cdot \xi^+)\} -\exp \{ - i \dt \q \cdot \xi \} \right) \d\n,
\end{multline*}
where
$$\xi^+=\xi-2 (\xi \cn) \n.$$
Since the last integral on the unit sphere only depends on $|\xi|$ and $\xi \cdot \q$, and $|\xi^+|=|\xi|$, we conclude that
$$\Qh = 0 \qquad \qquad \text{ for any } \qquad \xi \in \R.$$
We proved

\begin{theo}\label{existence}
The Maxwellian distribution
$$M(\v)=\left(\dfrac{m}{2\pi
T^{\#}}\right)^{3/2}\exp \{-\dfrac{m(\v-\u)^2}{2T^{\#}}\}  \qquad \v \in \R,$$
with $ T^{\#}=\dfrac{(1-\a)(1-\b)}{1-\a(1-\b)}T_1$ is an
equilibrium state for $\Q.$\end{theo}

\begin{nb}[Universality of the Maxwellian]\label{universal} One can easily generalize the above
computations to show that the Maxwellian \eqref{maxw} is an equilibrium state of any collision operator $\Qb$ taking the
weak form
\bq \label{qb}\langle \psi, \Qb(f) \rangle=\dpi \Itt B(\q,\n) f(\v)M_1(\w)[\psi(\vh)-\psi(\v)]\d\v\d\w\d\n,
\eq
for any smooth function $\psi.$ In \eqref{qb} the collision kernel $B(\cdot,\cdot)$ is assumed to be given by
\bq
\label{coll} B(\q,\n)=|\q|^\gamma b( \q \cn /|\q| )
\eq
 with $-1 \leq \gamma \leq 1$ and
$b(\cdot)$ nonnegative.
This shows that, as it happens in the elastic
case, \textit{the equilibrium state of the dissipative (linear)
Boltzmann equation is universal} in the sense that it does not
depend on the collision kernel.\end{nb}

\begin{nb} Note that for a general collision kernel $B(\q,\n)$ it
is convenient to use \eqref{qb} as a definition for $\Qb$ instead
of its strong form:
\bqs \Qb(f) =\dpi \It B(\q,\n)\left \{ J(\q,\n)
f(\vb)M_1(\wb)-f(\v)M_1(\w)\right\}\d\v\d\w\d\n\eqs
where the factor $J$ depends of $B,$ $\a$ and $\b$ in a
complicated way.\end{nb}

\section{On the trend to equilibrium}\label{H}

In this section we investigate the large--time behavior of the
solution  to the linear dissipative Boltzmann equation
\eqref{bolt1}. Precisely, let $f_0$ be a given (nonnegative)
distribution function and consider the Cauchy problem
\bq \label{cauch} \begin{cases} \dfrac{\partial }{\pt}f(\v,t)=\Q(f)(\v,t) \qquad
\qquad \v \in \R,\:t \geq 0\\
f(\v,t=0)=f_0(\v). \end{cases}\eq
Since the above problem is linear (and homogeneous), it is not
difficult to construct a (nonnegative) mild solution to
\eqref{cauch} by a simple iterative method. Moreover, this
solution is unique and preserves the mass
$$\IR f(\v,t)\d\v =\IR
f_0(\v)\d\v \qquad \text{ for any } t \geq 0.$$
 For further details, we refer  to \cite{petter} where the problem is studied in high generality, including the spatially
inhomogeneous equation with suitable boundary conditions.\medskip

A fundamental task in kinetic theory is to determine whether the solution to \eqref{cauch} converges toward the equilibrium
state of $\Q$ or not. Assuming the existence of a unique equilibrium distribution, this result has been proved by R.
Petterson \cite{petter} for collision kernels of the form \eqref{coll} with $-1 < \gamma < 1$ (corresponding to hard or soft
interactions).

Actually, the proof of \cite{petter} requires to prove that the equilibrium state we exhibited is unique. This can be done by
means of the $H$-theorem.\medskip

\noindent {\bf The $H$-theorem.} Let us recall here that the linear $H$-theorem for the dissipative Boltzmann equation
\eqref{bolt1} has been first established by R. Pettersson \cite{petter}, assuming the existence of an equilibrium state for
$\Q$. We point out that, in contrast to what happens in the nonlinear setting, the existence of such a steady--state is {\it
necessary} in order to prove the  $H$-theorem.

Once the existence of a steady state has been established, one can state the corresponding $H$-theorem and  to prove the
uniqueness of the stationary solution as a corollary.

We give here an elementary formal proof of the $H$-theorem (see
Pettersson \cite{petter} for a more general result for the linear
inhomogeneous equation with suitable boundary conditions). Let
$\Phi \::\:\mathbb{R}^+ \to \mathbb{R}$ be a {\it convex}
$C^1$--function. The associated entropy functional reads
\bq \label{Hphi} H_{\Phi}(f|M)=\int_{\R}M(\v) \Phi\left (\dfrac{f(\v)}{M(\v)} \right)\d\v , \eq
where $M(\v)$ is the Maxwellian \eqref{maxw}. The $H$-theorem
asserts that $H_{\Phi}(\cdot|M)$ is a Lyapunov functional for the
solution of the linear Boltzmann equation.

\begin{theo}[$H$-theorem] Let $\Phi \::\:\mathbb{R}^+ \to \mathbb{R}$ be a {\it convex}
$C^1$-function and let $f_0$ be a distribution function with unit mass such that $H_{\Phi}(f_0|M) < \infty.$ Then,
\begin{equation}\label{h}
\dfrac{d}{dt}H_{\Phi}(f(t)|M)\leq 0 \qquad \qquad (t \geq 0)
\end{equation}
where $f(t)$ stands for the (unique) solution to
\eqref{cauch}.\end{theo}

\begin{preuve} It is clear that
\bqs \begin{split}
\dfrac{d}{dt}H_{\Phi}(f(t)|M)&=\int_{\R}\dfrac{\partial
f}{\partial t}(\v,t)
\Phi'\left (\dfrac{f(\v,t)}{M(\v)} \right)\d\v\\
&=\int_{\R} \Q(f)(\v,t) \Phi'\left (\dfrac{f(\v,t)}{M(\v)}
\right)\d\v \end{split}\eqs
and the proof of \eqref{h} amounts to show that
\bq \label{entropie} \int_{\R}\Q(f)(\v)\Phi'\left
(\dfrac{f(\v)}{M(\v)} \right)\d\v \leq 0 \eq
for any distribution function $f$ with unit mass for which the
above integral is meaningful. From \eqref{weak}
\bqm \int_{\R}\Q(f)(\v)\Phi'\left (\dfrac{f(\v)}{M(\v)}
\right)\d\v= \dpi \Itt \qn f(\v)M_1(\w)\times \\
\times\left\{\Phi'\left (\dfrac{f(\vh)}{M(\vh)} \right)-
\Phi'\left(\dfrac{f(\v)}{M(\v)} \right) \right\} \d\v\d\w\d\n
\end{multline*}
and this last integral is also equal to
\begin{multline*}
\Itt \qn M(\v)M_1(\w) \left\{ \left [ \dfrac{f(\v)}{M(\v)} -
\dfrac{f(\vh)}{M(\vh)} \right] \Phi'\left (\dfrac{f(\vh)}{M(\vh)}
\right) \right.+ \\
\left. \dfrac{f(\vh)}{M(\vh)}  \Phi'\left (\dfrac{f(\vh)}{M(\vh)}
\right) - \dfrac{f(\v)}{M(\v)} \Phi'\left(\dfrac{f(\v)}{M(\v)}
\right) \right\} \d\v\d\w\d\n.\end{multline*}
Now, since $\langle \dfrac{f}{M}\Phi'\left(\dfrac{f}{M}\right),
\Q(M) \rangle=0$,
\begin{multline}\label{dissipation}
\int_{\R}\Q(f)(\v)\Phi'\left (\dfrac{f(\v)}{M(\v)} \right)\d\v=
\dpi \Itt \qn M(\v)M_1(\w) \times \\
\times \left\{\dfrac{f(\v)}{M(\v)} - \dfrac{f(\vh)}{M(\vh)}
\right\}\Phi'\left (\dfrac{f(\vh)}{M(\vh)} \right)\d\v\d\w\d\n.
\end{multline}
The conclusion follows since, $\Phi$ being convex,
$$\Phi'(a)(b-a) \leq \Phi(b)-\Phi(a) \qquad \qquad (a,\,b \in \mathbb{R})$$
and $\langle \Phi\left(\dfrac{f}{M}\right), \Q(M) \rangle=0.$
\end{preuve}

Direct consequence of the above result is the following uniqueness result (due to Pettersson \cite{petter} in a more general
setting).

\begin{cor} The Maxwellian distribution $M$ given by \eqref{maxw}
is the unique stationary solution to \eqref{bolt1} with unit mass.
\end{cor}

\begin{preuve} Let $F$ be another equilibrium state with unit
mass. Then, according to \eqref{dissipation} with
$\Phi(z)=\dfrac{(z-1)^2}{2}$ one sees that
$$\Itt \qn \,M(\v)M_1(\w) \left \{
\dfrac{F(\v)}{M(\v)}-\dfrac{F(\vh)}{M(\vh)} \right \}
\dfrac{F(\vh)}{M(\vh)}\d\v\d\w\d\n=0.$$
This implies that
\begin{multline*}
\Itt \qn M(\v)M_1(\w) \left \{
\dfrac{F(\v)}{M(\v)}-\dfrac{F(\vh)}{M(\vh)} \right \}^2
\d\v\d\w\d\n =\\
=\Itt \qn F(\v)M_1(\w) \left \{
\dfrac{F(\v)}{M(\v)}-\dfrac{F(\vh)}{M(\vh)} \right \}
\d\v\d\w\d\n=0\end{multline*} where this last integral is null
since $\Q(F)=0.$ Consequently, one gets that
$$\dfrac{F(\v)}{M(\v)}=\dfrac{F(\vh)}{M(\vh)} \qquad \qquad \text{
for any } \quad (\v,\w) \in \R$$ and this last identity leads, as
in the elastic case, to $F=M.$\end{preuve}

\noindent {\bf The evolution of the second moment.} To prove that the solution to the Cauchy problem \eqref{cauch} converges
towards the (unique) equilibrium state, one has to establish some  {\it a priori} estimates on moments. Actually, in contrast
to the elastic case and because of the lack of collision invariants, it is not trivial to estimate the evolution of the
moments of $f(\v,t)$. This difficulty is peculiar to the hard--spheres model and does not occurs for pseudo--Maxwellian
molecules \cite{spiga}, since in this case the equations for moments are in close form. For long--range interactions forces,
R. Pettersson succeeded in proving uniform estimates on the higher moments of $f(\v,t)$ (see \cite[Theorem 4.1]{petter}) for
various types of kernels. Unlikely, his arguments do not apply to the hard--sphere model. Here we  show that the second
moment of the solution to \eqref{cauch} remains uniformly bounded in time. Precisely, let the initial distribution function
$f_0 \in L^1(\R)$ have \textit{unit mass} and let $f(t)$ be the solution to the Cauchy problem \eqref{cauch}. Recall that for
any $t \geq 0$, $f(\v,t)$ has also unit mass. One considers the following second moment of $f(t)$:
$$T(t)=\dfrac{m}{3} \IR f(\v,t) (\v-\u)^2\d\v.$$
Note that $T(t)$ is not {\it stricto sensu} the temperature of $f(t)$ which is defined by replacing the velocity $\u$ by the
drift velocity of $f(t)$. Define also
$$F(t)=\IRR |\v-\w|^2 f(\v,t)M_1(\w)\d\v\d\w.$$
One has
\begin{multline*}
F(t)=\IR (\v-\u)^2 f(\v,t)\IR M_1(\w)\d\w + \IR \ft \d\v \IR
(\w-\u)^2M_1(\w)\d\w\\
-2 \IR (\v-\u)\ft \d\v \cdot \IR (\w-\u)M_1(\w)\d\w .
\end{multline*}
The last integral is equal to zero by definition of $\u.$ Therefore
\bq \label{F-T} F(t)=\dfrac{3}{m}T(t) + \dfrac{3}{m_1}T_1.\eq
Now, from \eqref{weak}, one has
$$
\dfrac{\d \,T(t)}{\d t}=\dfrac{m}{6\pi}\Itt \qn \ft M_1(\w) \left
\{(\vh-\u)^2-(\v-\u)^2 \right\}\d\v\d\w\d\n
$$
and
\bqs \begin{split} (\vh-\u)^2-(\v-\u)^2 &=4 \a^2 (1-\b)^2 \qn ^2
-4 \a (1-\b) (\q \cn) (v-\u) \cn\\
&= -4 \a (1-\b) [1-\a(1-\b)]\, \qn^2 + 4 \a (1-\b) (\q \cn) (\w
-\u) \cn. \end{split}\eqs
Consequently,
\begin{multline*}
\dfrac{\d \,T(t)}{\d t}=-\dfrac{2m}{3\pi} \mu (1-\a(1-\b))
\Itt \qn ^3 \ft M_1(\w)\d\v\d\w\d\n \\
+ \dfrac{2m}{3\pi} \mu \Itt \qn (\q \cn) ((\w-\u) \cn) \ft M_1(\w)\d\v\d\w\d\n.
 \end{multline*}
Recall that $0< \mu=\a (1-\b)< 1$. Since
$$\IS \qn ^3 \d\n=\pi |\q|^3 \qquad \text{ and } \qquad \IS \qn (\q \cn) (\w-\u)
\cn\,\d\n=\pi |\q| \q \cdot (\w-\u),$$
one gets
\begin{multline*}
\dfrac{\d \,T(t)}{\d t} \leq -\dfrac{2m}{3}\mu (1-\mu)
\It |\q|^3 \ft M_1(\w)\d\v\d\w \\
+ \dfrac{2m}{3} \mu \It |\q|^2 |\w-\u| \ft M_1(\w)\d\v\d\w.
\end{multline*}
Let us first look for a upper bound to the second integral in terms of $F(t)$. Using the fact that $\displaystyle \IR
(\w-\u)|\w-\u|\,M_1(\w)\d\w=0$, we obtain
\begin{multline*}
\It |\q|^2 |\w-\u| \ft M_1(\w)\d\v\d\w =\IR (\v-\u)^2 f(\v,t)\IR |\w-\u|\,M_1(\w)\d\w \\+ \IR \ft \d\v \IR
|\w-\u|^3M_1(\w)\d\w.
\end{multline*}
 Thus,
\bq \label{estim1} \It |\q|^2 |\w-\u| \ft M_1(\w)\d\v\d\w \leq C_1 F(t), \eq
where
$$
C_1=\max \left\{ \IR |\w-\u|M_1(\w)\d\w\:,\;\dfrac{\IR
|\w-\u|^3M_1(\w) \d\w}{\IR |\w-\u|^2M_1(\w) \d \w} \right\}
$$
is a positive (explicit) constant depending only on $M_1.$
Moreover, Jensen's inequality gives
\bq \label{estim2} \IRR |\q|^3 \ft M_1(\w)\d\v\d\w \geq \left( \IRR |\q|^2 \ft M_1(\w)\d\v\d\w\right)^{3/2}. \eq
Now, combining \eqref{estim1} and \eqref{estim2} and \eqref{F-T}
one gets the differential inequality
\bq \label{ine1}
\dfrac{\d \,F(t)}{\d t} \leq -2\mu (1-\mu) F(t)^{3/2} + 2\mu C_1 F(t).
\eq
A direct inspection then shows that  the solution  to \eqref{ine1} satisfies the bound
$$F(t) \leq \max \left\{ \dfrac{C_1^2}{(1- \mu)^2}, F(0)
\right\}\qquad \qquad t \geq 0.$$ Turning back to $T(t)$ one
obtains that
\bq \label{boundT} \sup_{t \geq 0} \IR (\v-\u)^2 \ft \d\v < \infty
\eq
provided $\displaystyle \IR \v^2 f_0(\v)\d\v < \infty$. Note
that the above bound for $T(t)$ is explicitly computable
in terms of $f_0$, $C_1$, $\a$ and $\b$.
\medskip

Now, mass conservation and the $H$-theorem, together with estimate
\eqref{boundT} imply that, provided
\bq \label{estimatef0} \IR \left(1 + \v^2 + |\log f_0(\v)|\right)
f_0(\v) \d\v \le K < \infty \eq
at any subsequent time $t >0$
\bqs \IR \left( 1 + (\v-\u)^2 + |\log \ft|\right )
\ft \d\v \le K_1 < \infty ,\eqs
and this implies the weak--compactness in $L^1(\R)$ of the family
$\{f(\v,t)\}_{t \geq 0}$. Now, following the strategy of \cite{petter}, one obtains first
the weak--convergence,  and
using translation continuity,  the strong
convergence towards the equilibrium of $\ft$.
\begin{theo}
Let $f_0 \in L^1(\R)$ be a distribution function with unit mass
satisfying \eqref{estimatef0} and let $f(\v,t)$ be the solution to
the Cauchy problem \eqref{cauch}. Then
$$\lim_{t \to \infty}\|f(t)-f_0\|_{L^1(\R)}=0.$$
\end{theo}

\begin{nb} We conjecture that, as it occurs for the pseudo-Maxwellian
approximation \cite{spiga}, the decay of $\|f(t)-f_0\|_{L^1(\R)}$
towards $0$ is exponential (with an explicit rate).\end{nb}

\section{Appendix}\label{appen}

Let us make rigorous the derivation of the Fokker--Planck equation
we formally obtained in Section \ref{graz}. This can be done {\it
a posteriori} using the results of the previous Section. We maintain here the notations
 of Section \ref{graz}. The only difference is that,
hereafter $\Qe$ denotes the collision operator with kernel
$$B_{\var}(\q,\n)=b_{\var}(\t)\qn ,$$
whereas in Section \ref{graz} we considered a re-scaled kernel
(see Remark \ref{scale}). Actually, in Section \ref{graz} we were
concerned with an approximation of the \textit{Boltzmann collision
operator} whereas in this Appendix, we approximate the
\textit{solution to the Boltzmann equation} as collisions become
grazing. This difference will appear clearly in the sequel. Let
$f_0$ be a nonnegative distribution function fulfilling estimate
\eqref{estimatef0}
and
consider the following Cauchy problem
\bq \label{cauchyeps}
 \begin{cases}
\dfrac{\partial f_{\var}}{\partial t}(\v,t)=\Qe(f_{\var})(\v,t)
\qquad \qquad t > 0, \:\v \in \R\\
f_{\var}(\v,0)=f_0(\v).\end{cases}\eq
As we outlined in  Section \ref{H}, problem
\eqref{cauchyeps} admits a (unique) weak-solution which satisfies
\begin{multline}\label{weakt}
-\inrt \IR \fd \partial_t \psi(\v,t) \d \v  - \IR f_0(\v) \psi(\v,0)\d \v \\
=\dpi \inrt \Itt B_{\var}(\q,\n) \fd M_1(\w) \left[
\psi(\vh,t)-\psi(\v,t)\right] \d \v \d \w \d \n ,\end{multline}
for any $\psi \in C^1_{2,c}([0,\,+\infty[ \times \R)$. In usual notations, $C^1_{2,c}([0,\,+\infty[ \times \R)$ denotes
the space of all functions
$\psi$ which are continuously differentiable with compact support in
$[0,\,+\infty[$ and twice continuously differentiable in $\R.$
We recall that
\bq I_{\var}=\int_0^{\pi/2}b_{\var}(\t)\cos^3 \t\, \sin \t \,\d\t
\sim \dfrac{\var^4}{2\pi} \qquad \text{ as } \qquad \var \sim
0.\eq
Let us introduce the time--scaling
$$g_{\var}(\v,t)=f_{\var}(\v,\var^{-4}\,t).$$
Then, considering a test-function of the form
$\psi_{\var}(\v,t)=\psi(\v,\var^{-4}\,t)$ into \eqref{weakt} leads
to
\begin{multline}\label{weakt2}
-\inrt \IR g_{\var}(\v,t) \partial_t \psi(\v,t) \d \v  - \IR f_0(\v) \psi(\v,0)\d \v \\
= \dfrac{1}{2\pi \var^{4}} \inrt \Itt B_{\var}(\q,\n)
g_{\var}(\v,t) M_1(\w) \left[ \psi(\vh,t)-\psi(\v,t)\right] \d \v
\d \w \d \n.\end{multline}
The key point of the approximation procedure is the following
\begin{propo}\label{wconv} There exists a nonnegative function $g\::\:[0,\,+\infty[ \to L^1(\R)$
and a subsequence, still denoted $(g_{\var})_{\var \geq 0}$ such
that $g_{\var}$ converges weakly in $L^1_{loc}([0,\,+\infty[,
L^1(\R))$ towards $g$ as $\var$ goes to zero.\end{propo}
We leave the proof to Proposition \ref{wconv} to the end of this
Appendix and explain now how to derive the Fokker--Plank equation
from it. Inserting the expansion \eqref{taylor} into
\eqref{weakt2} leads to the approximation
\begin{multline*}
-\inrt \IR g_{\var}(\v,t) \partial_t \psi(\v,t) \d \v  - \IR
f_0(\v) \psi(\v,0)\d \v = \int_0^{\infty} \mathbf{J}^1_{\var}(t)
\d t + \int_0^{\infty}
\mathbf{J}^2_{\var}(t) \d t + \mathbf{R}_{\var}\\
= - \dfrac{\z}{ 2\pi \var^4} \inrt \Itt B_{\var}(\q,\n)(\q \cn)
g_{\var}(\v,t) M_1(\w) \nabla_\v
\f(\v,t) \cn \,\d\v\d\w\d\n\\
+ \dfrac{(\z )^2}{4\pi \var^4} \inrt \Itt B_{\var}(\q,\n)\left(\q
\cn \right)^2 g_{\var}(\v,t) M_1(\w) \mathbb{D}^2\f(\v,t) \cdot \n
\otimes \n
\,\d\v\d\w\d\n + \mathbf{R}_{\var}.\\
\end{multline*}
In the above expression, $\mathbf{R}_{\var}=\mathbf{R}_{\var}(\psi)$ is a
remainder term obtained from the Taylor expansion of $\psi$
\eqref{taylor} (see \cite{goudon} for details). One can prove as
in the elastic case \cite{goudon} that
\bq \label{Rpsi}\lim_{\var \to 0} \mathbf{R}_{\var}(\psi)=0 \eq
for any test--function $\psi \in C^1_{2,c}([0,\,+\infty[ \times
\R)$. As in Section \ref{graz} one can show that
$$\int_0^{\infty}\mathbf{J}^1_{\var}(t) \d t=-I_{\var}\dfrac{\z}{\var^4} \inrt \IRR |\q|^2
g_{\var}(\v,t) M_1(\w) \nabla_\v \f(\v,t) \cdot \dfrac{\q}{|\q|}
\,\d\v\d\w.$$
Consequently, thanks to Proposition \ref{wconv}
\bq \label{limj1t} \lim_{\var \to
0}\int_0^{\infty}\mathbf{J}^1_{\var}(t)\,\d
t=-\frac{\a(1-\b)}{\pi} \inrt \IRR |\q|^2 g(\v,t) M_1(\w)
\nabla_\v \f(\v,t) \cdot \dfrac{\q}{|\q|} \,\d\v\d\w.\eq
We proceed in the same way for $\mathbf{J}^2_{\var}$ to obtain
\bq \label{limj2t} \lim_{\var \to
0}\int_0^{\infty}\mathbf{J}^2_{\var}(t) \d t=
\dfrac{\a^2(1-\b)^2}{2\pi} \inrt \IRR |\q|^3 g(\v,t) M_1(\w)
\mathbb{D}^2\f(\v,t) \cdot \mathbf{Diag}[0,1,1]\,\d\v\d\w.\eq
Then, combining \eqref{weakt2}, \eqref{limj1t} and \eqref{limj2t},
one sees that the weak limit $g$ satisfies
\begin{multline}\label{weakform} -\inrt \IR g(\v,t)
\partial_t \psi(\v,t) \d \v  - \IR f_0(\v) \psi(\v,0)\d \v=
 - \dpi \inrt \IR  \d\v \\
 \nabla_{\v}\psi(\v,t)\cdot
 \IR |\v-\w|^3 \mathbb{S}(\v-\w)
\left\{\k M_1(\w)\nabla_{\v}g(\v,t) + (\k -\mu)
g(\v,t)\nabla_{\w}M_1(\w)\right\}\d\w ,
\end{multline}
where $\k$ and $\mu$ are defined in Section \ref{graz}. At this point it is not
difficult to recognize in \eqref{weakform} the weak
formulation of the Cauchy problem
\bq \label{eqFP}\begin{cases} \dfrac{\partial g}{\partial
t}(\v,t)=\dpi \QF (g)(\v,t) \qquad \qquad t \geq 0, \v \in \R \\
g(\v,0)=f_0(\v)\end{cases} \eq
where the Fokker--Planck collision operator is given by
\eqref{FP}. It is well--known that problem \eqref{eqFP} admits a
(unique) nonnegative weak solution $g$. We proved

\begin{theo}
There exists a subsequence, still denoted $g_{\var}$, such that
$$g_{\var} \underset{\var \to 0}{\rightharpoonup} g \text{ weakly  in }
L^1_{loc}([0,+\infty[,\,L^1(\R))$$ where $g$ is a weak solution to
the Fokker--Planck equation \eqref{eqFP} with initial datum $f_0$
satisfying \eqref{estimatef0}.
\end{theo}
It remains now to prove Proposition \ref{wconv}. Clearly, it is
enough to prove that the sequence $g_{\var}$ satisfies the  uniform estimate
\bq \sup_{\var \geq 0, t \geq 0} \IR (1+(\v-\u)^2 + |\log
g_{\var}(\v,t)|)g_{\var}(\v,t) \d \v < \infty.\eq
Now the estimate
$$\sup_{\var \geq 0, t \geq 0} \IR (1+ |\log
g_{\var}(\v,t)|)g_{\var}(\v,t) \d \v < \infty$$
follows from the $H$-theorem applied to $f_{\var}(t)$. Note that
the $H$-theorem turns to be valid for the collision kernel
$B_{\var}(\cdot)$ since the equilibrium state is universal (see
Remark \ref{universal}). Now to prove the remaining estimate, we
proceed as we did in Section \ref{H} to derive formula
\eqref{boundT}. We only sketch here the main changes. Let
$$T_{\var}(t)=\dfrac{m}{3} \IR g_{\var}(\v,t) (\v-\u)^2\d\v=\dfrac{m}{3} \IR f_{\var}(\v,\var^{-4}t) (\v-\u)^2\d\v.$$
Then,
$$
\dfrac{\d \,T_{\var}(t)}{\d t}=\dfrac{m \var^4}{6\pi}\Itt
B_{\var}(\q \cn) f_{\var}(\v, \var^{-4}t) M_1(\w) \left
\{(\vh-\u)^2-(\v-\u)^2 \right\}\d\v\d\w\d\n.
$$
Since
%
$$\IS \qn ^2 B_{\var}(\q \cn) \d\n=2\pi I_{\var} |\q|^3, $$

 $$\IS B_{\var}(\q \cn) (\q \cn) (\w-\u)
\cn\,\d\n \leq 2\pi I_{\var} |\q|^2 |\w-\u|$$
and  $\lim_{\var  \to 0} 2\pi \var^4 I_{\var}=1$,
we can draw the same conclusion of Section \ref{H}.

\section{Conclusions}

We proved existence and uniqueness of a collision equilibrium
for the dissipative linear Boltzmann equation with general
collision kernel. This equilibrium state is a \textit{universal
Maxwellian} with the same mass velocity as the field particles
background and with a (non--zero) temperature always lower than
the one of the background (depending on mass ratio and
inelasticity). As early noticed in \cite{spiga},   this follows
from the combined effects of momentum and energy exchange with
fields particles on the one side and, on the other side, of energy
dissipation in the binary collisions.

We point out that the existence of a Maxwellian equilibrium at non--zero temperature is of primary importance  to reckon the
hydrodynamic equations for the considered granular flow. This can be done through  a  Chapman--Enskog procedure (see the
conclusions of \cite{spiga}).

In space homogeneous conditions, we showed that the solution to
the linear dissipative Boltzmann equation converges towards the
equilibrium state as time goes to infinity for any initial datum
with finite entropy and temperature. Unlikely, our convergence
result is based upon compactness arguments and we have not been
able to determine the decay rate towards the equilibrium. We may
hope that, as it occurs for the pseudo--Maxwellian approximation
\cite{spiga}, the relaxation to equilibrium is exponential. The
results of  Appendix show that the Fokker--Planck equation
\eqref{eqFP} is a good approximation of \eqref{bolt1} when
collisions become grazing. It is well--known (see \cite{fokker})
that the solution to \eqref{eqFP} relaxes to $M$ exponentially
with an explicit rate related to $T^{\#}$. This supports us in the
belief that the same occurs for the dissipative Boltzmann equation
\eqref{bolt1}. We may infer that dissipation--dissipation entropy
methods should lead to such a result. Work is in progress in this
direction.

\par
\bigskip
\bigskip

\par

\noindent {\bf Acknowledgement:} This research was carried out while B. Lods was enjoying a post--doctoral position at the
IMATI--CNR/Department of Mathematics of the University of Pavia. He would like to express his sincere gratitude to P. Pietra
and Professor G. Toscani for their kind hospitality during this period. B.L. is supported by the IHP project ``HYperbolic and
Kinetic Equations'', No.~HPRN-CT-2002-00282, funded by the EC. G.T. acknowledges financial supports both from the project
``HYperbolic and Kinetic Equations'', funded by the EC., and from the Italian MURST, project ``Mathematical Problems in
Kinetic Theories''.

\end{document}